# Quantum oscillation studies of topological semimetal candidate ZrGe*M* (*M* = S, Se, Te)

J. Hu[1*], Y.L. Zhu[1], D. Graf[2], Z.J. Tang[1], J.Y. Liu[1], and Z.Q. Mao[1*]

[1]Department of Physics and Engineering Physics, Tulane University, New Orleans, LA 70118, USA

[2]National High Magnetic Field Lab, Tallahassee, FL 32310, USA

Abstract

The *WHM* - type materials (*W*=Zr/Hf/La, *H*=Si/Ge/Sn/Sb, *M*=O/S/Se/Te) have been predicted to be a large pool of topological materials. These materials allow for fine tuning of spin-orbit coupling, lattice constant and structural dimensionality for various combinations of *W*, *H* and *M* elements, thus providing an excellent platform to study how these parameters' tuning affect topological semimetal state. In this work, we report the high field quantum oscillation studies on ZrGe*M* (*M*=S, Se, and Te). We have found the first experimental evidence for their theoretically-predicted topological semimetal states. From the angular dependence of quantum oscillation frequency, we have also studied the Fermi surface topologies of these materials. Moreover, we have compared Dirac electron behavior between the ZrGe*M* and ZrSi*M* systems, which reveals deep insights to the tuning of Dirac state by spin-orbit coupling and lattice constants in the *WHM* system.

[*]jhu@tulane.edu; zmao@tulane.edu

The discoveries of topological semimetals have opened a new era of condensed matter physics. These materials represent new topological states of quantum matter and exhibit exotic properties resulting from relativistic fermions hosted by Dirac or Weyl cones, such as extremely high bulk carrier mobility [1-7], large magnetoresistance [1-7], and potential topological superconductivity [8], which hold tremendous potential for technology applications. In three dimensional (3D) Dirac semimetals such as $Na_3Bi$ [9-10] and $Cd_3As_2$ [11-14], the four-fold degenerate band crossings at Dirac nodes are protected by the crystal symmetry. When the spin degeneracy is lifted by broken time-reversal or spatial inversion symmetry, a Dirac state evolves to a Weyl state where each Dirac cone splits to a pair of Weyl cones with opposite chirality[9, 11, 15-16]. The inversion symmetry broken Weyl state has been demonstrated in transition metal monopnictides (Ta/Nb)(As/P) [15-22], photonic crystals [23], and $(W/Mo)Te_2$ [24-33]. The spontaneous time reversal symmetry breaking Weyl state has been reported in $YbMnBi_2$ [34] and predicted in magnetic Heusler alloys [35-39] and the magnetic members of R-Al-X (R=rare earth, X=Si, Ge) compounds [40].

In addition to the above topological semimetals with discrete Dirac/Weyl nodes in momentum space, another type of topological semimetal - the topological nodal line semimetal which features Dirac bands crossing along a one-dimensional line/loop, has also been predicted [41-47] and experimentally observed in several material systems such as $(Pb/Tl)TaSe_2$ [48-49], ZrSi*M* (*M*=S, Se,Te) [50-54] and $PtSn_4$ [55]. Among these materials, the layered material ZrSi*M* exhibits unique properties. It hosts two types of Dirac states [50], *i.e.* the three dimensional (3D) nodal-line state with an exceptionally wide energy range of linear dispersion, and the two dimensional (2D) Dirac state protected by the non-symmorphic symmetry. These two types of Dirac states were experimentally demonstrated by APRES measurements [50-51] and quantum oscillation experiments for ZrSiS [56-59]. Replacing S by Se and Te greatly reduces the interlayer binding energy, but preserves the main topological signatures in the electronic structure [46]. Indeed, the topological Dirac nodal-line states in ZrSiSe and ZrSiTe have been experimentally demonstrated by quantum oscillation studies [52] and later confirmed by ARPES observations [53-54]. Furthermore, due to reduced interlayer binding, atomically thin 2D layers of ZrSiSe and ZrSiTe can be obtained by mechanical exfoliation (see supplement materials in [52]), which opens up the possibility to realize the predicted 2D topological insulator in monolayers [46], and provides a new platform for the investigation of new topological fermion physics in low dimensions.

ZrSi*M* compounds belong to a larger family of materials *WHM* with the PbFCl-type structure (*W*=Zr/Hf/La, *H*=Si/Ge/Sn/Sb, *M*=O/S/Se/Te) [46]. For the compounds with different combinations of *W*, *H* and *M* elements, their overall electronic structures are predicted to be similar and display nodal-line states, besides some small discrepancies caused by the variations of spin-orbit coupling (SOC) strength and structural dimensionality [46]. Indeed, in addition to nodal-line states in the aforementioned ZrSi*M* [50-52, 60], topologically non-trivial phases have also been observed in other *WHM*-type materials such as ZrSnTe [61] and HfSiS [62-64]. Given that the SOC strength and the structural dimensionality governed by the steric-electronic balance can be fine-tuned with different combinations of *W*, *H* and *M* elements, *WHM* compounds provide an ideal platform to investigate the evolution of the topological fermion properties with these parameters. Therefore, experimental demonstrations and characterizations of the predicted topological states in other *WHM* compounds are important.

With the above motivation, we have extended our studies to the ZrGe*M* (*M*=S, Se,Te) system. Similar to other *WHM* compounds [46], ZrGe*M* also possesses a layered tetragonal structure formed from the stacking of M-Zr-Ge-Zr-M slabs (Fig. 1a) [65-66]. Compared to ZrSi*M*, ZrGe*M* has an enhanced interlayer binding energy [46]; hence its electronic structure is expected to be more 3D. Although the first principle calculations have shown ZrGe*M* shares similar electronic structures near the Fermi level with ZrSi*M* [46], *i.e.* hosting topological Dirac nodal-line state, experimental verification has not been reported thus far. In this work, we report the quantum oscillation studies on ZrGe*M* compounds and show the first experimental evidence for topological fermions in this family of materials.

The ZrGe*M* single crystals (Fig. 1c) were synthesized by a chemical vapor transport method similar to that used for growing ZrSi*M* single crystals [52, 56]. The compositions of the synthesized crystals were analyzed using Energy dispersive spectroscopy (EDS). The excellent crystallinity is demonstrated by the sharp (00L) x-ray diffraction peaks, as shown in Fig. 1b. Due to the increased ionic radius from S to Te ions, the *c*-axis lattice parameter, *i.e.*, the M-Zr-Ge-Zr-M slab thickness, is increased from 8.041 Å for ZrGeS to 8.305 Å for ZrGeSe and to 8.653 Å for ZrGeTe. A similar trend was also observed in the ZrSi*M* compounds [52].

Evidence for topological Dirac states in ZrGe*M*, including light effective mass, high mobility, and non-trivial Berry phase, have been found in our quantum oscillation experiments, which were performed using the 31T resistive magnet in NHMFL, Tallahassee. In our measurements, both Shubnikov–de Haas (SdH) and de Haas-van Alphen (dHvA) oscillations have been observed in ZrGe*M*. However, the SdH oscillations are much weaker than the dHvA oscillations (shown later). Our previous work on ZrSiS has shown the dHvA effect better reveals intrinsic Dirac fermion properties [56]. Therefore, our analyses will be mainly focused on the dHvA effect, which was probed in magnetic torque measurements using a piezoresistive cantilever. Given the torque signal is expected to vanish when magnetic field is perfectly aligned along the out-of-plane (*B*//*c*) and in-plane (*B*//*ab*) directions, we performed the measurements with the fields nearly along the out-of-plane and in-plane directions (denoted by *c'* and *ab'* respectively). In Figs. 2a and 2d, we have presented the field dependences of magnetic torque at different temperatures for ZrGeS for *B*//*c'* and *B*//*ab'*, respectively. For both field orientations, we observed strong dHvA oscillations at low temperatures. The oscillations vanished when the temperature was increased above 30K. The presence of quantum oscillations for both field orientations indicates a 3D Fermi surface in ZrGeS despite its layered crystal structure, which will be discussed in more details later. From the oscillatory torque $\Delta\tau$ obtained by subtracting the smooth background (Figs. 2b and 2e), one can find that both oscillation patterns contain multiple frequency components. This can be clearly seen in the fast Fourier transform (FFT) analyses as shown in Figs. 2c and 2f. The dHvA oscillations are composed of one lower frequency $F_\alpha$ (=12.5T) and two higher frequencies $F_{\beta 1}$ (=236T) and $F_{\beta 2}$ (=380T) for *B*//*c'*, and one low frequency $F_\alpha$ =(17T) and one high frequency $F_{\beta 1}$ (=132T) for *B*//*ab'*. The coexistence of lower and higher frequencies have also been observed in ZrSiS [52, 57-58] and they are attributed to the 2D non-symmorphic Dirac state and the 3D nodal-line state, respectively [52].

For $B//c'$, we have also observed splitting of oscillation peaks for the lower- and higher- frequency components at low temperature, as indicated by the black and purple arrows in Fig. 2a, respectively. The split peaks gradually merge upon increasing temperature (Fig. 2a), which is a typical signature of Zeeman effect due to the broadening and overlapping of the split Landau levels at higher temperatures. Although our FFT analyses revealed multiple frequencies, from the careful inspection of the oscillation patterns (Fig. 2a), we can see that Zeeman splitting occurs for the $F_\alpha$ = 12.5T and $F_{\beta 1}$ =236T oscillation components. For the $F_\alpha$ component, due to the extremely low frequency and strong peak splitting, Zeeman effect could easily be overlooked and the split peaks could be mistakenly attributed to the normal oscillation peaks from unsplit LLs. Fortunately, its clear signatures of Zeeman effect can be resolved in susceptibility oscillations (d$M$/d$B$), which are obtained by taking the derivative of the lower frequency component due to $\tau \propto M$, as shown in the inset of Fig. 2b. From the peak splitting and the effective electron mass (see below), we have evaluated the $g$-factor of quasi-particles using a method discussed in ref. [56]; $g$ = 15 and 12, respectively, for the quasi-particles hosted by the $F_\alpha$ and $F_{\beta 1}$ bands, much larger than the $g$-factor of a free electron ($g \sim 2$). This result is consistent with our previous observation of strong Zeeman splitting with a large $g$-factor (~38) in ZrSiS [56].

More information about the Dirac fermions properties in ZrGeS can be extracted from the analyses of dHvA oscillations. The dHvA oscillations for a 3D system can be described by the 3D Lifshitz-Kosevich (LK) formula [67-68] with a Berry phase being taken into account for a topological system [69]:

$$\Delta\tau \propto -B^{1/2} R_T R_D R_S \sin[2\pi(\frac{F}{B}+\gamma-\delta)] \qquad (1)$$

where $R_T = \alpha T\mu/B\sinh(\alpha T\mu/B)$, $R_D = \exp(-\alpha T_D\mu/B)$, and $R_S = \cos(\pi g\mu/2)$. $\mu = m^*/m_0$ is the ratio of effective cyclotron mass $m^*$ to free electron mass $m_0$. $T_D$ is Dingle temperature, and $\alpha = (2\pi^2 k_B m_0)/(\hbar e)$. The oscillations of $\Delta\tau$ is described by the sine term with a phase factor $\gamma - \delta$, in which $\gamma = \frac{1}{2} - \frac{\phi_B}{2\pi}$ and $\phi_B$ is Berry phase. The phase shift $\delta$ is determined by the dimensionality of the Fermi surface and has a value of $\pm 1/8$ for 3D cases, with the sign depending on whether the probed extreme cross-section area of the FS is maximal (-) or minimal (+) [67]. From the LK formula, the effective mass $m^*$ can be obtained through the fit of the temperature dependence of the oscillation amplitude to the thermal damping factor $R_T$. In the case of multi-frequency oscillations, the oscillation amplitude for each frequency can be represented by the amplitude of FFT peak, and the parameter $1/B$ in $R_T$ should be the average inverse field $1/\bar{B}$, defined as $1/\bar{B} = (1/B_{max} + 1/B_{min})/2$, where $B_{max}$ and $B_{min}$ define the magnetic field range used for FFT. As shown in the insets of Figs. 1c and 1f, for all probed oscillation frequencies, the obtained effective masses are in the range of $0.05$-$0.1 m_0$ (see Table 1), which are only slightly larger than those obtained from the dHvA oscillation studies on ZrSiS [56], and agrees with the nature of massless relativistic fermions.

High quantum mobility and π Berry phase are also important characteristics of topological fermions. For the multi-frequency oscillations seen in ZrGeS, these parameters cannot be directly obtained through the conventional approaches, *i.e.*, the Dingle plot and the Landau Level fan diagram, but can be extracted through the fit of the oscillation pattern to the generalized multiband LK formula [70]. This method has been shown to be efficient for the analyses of the multi-frequency quantum oscillations in ZrSi*M* [52, 56]. The LK fit to the $T = 1.8$K oscillation pattern is very difficult for the case of $B//c'$ due to the presence of a strong Zeeman effect, but much easier for $B//ab'$, as shown in the inset of Fig. 2e. As summarized in Table I, from the LK fit for $B//ab'$, we have obtained Dingle temperatures of 28 and 35K, respectively, for the lower (17T) and higher frequencies (132T). The quantum relaxation time $\tau_q$ [$= \hbar/(2\pi k_B T_D)$] corresponding to such values of Dingle temperature are $4.4\times10^{-14}$ s and $3.5\times10^{-14}$ s, from which the quantum mobility $\mu_q$ ($= e\tau/m^*$) are estimated to be 1547 cm$^2$/Vs and 684 cm$^2$/Vs, respectively. The obtained quantum mobility values for ZrGeS are remarkably smaller than those of ZrSiS (2000-10000 cm$^2$/Vs [56]). Given that high mobility is the generic feature of topological relativistic fermions [1], the low quantum mobility seen in ZrGeS may imply that replacing Si by Ge results in some changes in Dirac crossings in band structure, which will be discussed in more details below. The relativistic nature of carriers in ZrGeS is demonstrated by the non-trivial Berry phases. From the LK-fit, we obtained the phase factors ($\gamma-\delta$) of 0.69 and 0.61 for the $F_\alpha$ and $F_{\beta l}$ bands respectively, from which the Berry phases are found be non-trivial, *i.e.* $(-0.38\pm0.25)\pi$ and $(-0.22\pm0.25)\pi$ (Table1). Such results are consistent with the theoretically predicted topological Dirac states in all *WHM* systems [46].

Similar features of topological fermions have also been seen in the isostructural compounds ZrGeSe and ZrGeTe. We also observed strong dHvA oscillations in magnetic torque measurements on ZrGeSe (Figs. 3a and 3d) and ZrGeTe (Figs. 4a and 4d). Like ZrGeS, ZrGeSe also exhibits dHvA oscillations with multiple frequencies for both $B//c'$ (Figs. 3b) and $B//ab'$ (Figs. 3e), which are clearly resolved in FFT spectra (Figs. 3c and 3f). In the FFT spectra, a magnetic breakdown effect is observed. This effect is caused by quantum tunneling of electrons between the different orbits on the different parts of Fermi surface [68], leading to additional frequencies equal to the sum or difference of fundamental frequencies, as shown in Figs. 3c and 3f. In addition to the magnetic breakdown, Zeeman splitting also appears at high field for both $B//c'$ and $B//ab'$, manifesting itself in peak splitting as shown in the inset of Fig. 3a and indicated by the arrows in Fig. 3d. In contrast with ZrGeS which displays strong splitting for both lower- and higher- frequency components when $B//c'$ (Fig. 2a), Zeeman splitting in ZrGeSe can be observed for both $B//c'$ and $B//ab'$ and is resolvable only for the higher- frequency component. Due to too many oscillation frequencies in ZrGeSe, we are unable to tell which frequency components exhibit Zeeman splitting, hence the values of $g$-factor cannot be determined. Another interesting feature in ZrGeSe is the inverse sawtooth-like oscillation pattern at low temperatures for $B//c'$ (Fig. 3a, inset), which is suggestive of a 2D/quasi-2D electronic state in some cases [68, 71]. Nevertheless, we note that inverse sawtooth-like oscillation pattern could also be caused by the torque interaction, which is essentially an instrumental effect due to the 'feedback' of the oscillating magnetic moment on the cantilever position [72]. Our observed inverse sawtooth-like oscillations most likely have this origin, since we did not observe such inverse sawtooth-like oscillations in ZrGeTe, whose electronic structure should have lower dimensionality than ZrGeSe as discussed below.

For the major fundamental frequencies seen in FFT spectra (Figs. 3c and 3f), from the fit of the temperature dependence of FFT amplitude (insets of Figs. 3c and 3f), the effective masses are estimated to be ~0.05-0.22 $m_0$ (Table II), larger than those for ZrGeS (Table I). Through the multiband LK-fit, we can also obtain the quantum mobility and Berry phase of quasi-particles in ZrGeSe. For $B//c'$, the fit was limited to the magnetic field range below $B$=20T (Fig. 3b, inset), to avoid the complexity induced by strong Zeeman splitting at high fields. For $B//ab'$, however, the fitting was not successful due to the complicated oscillation pattern. From the multi-band LK fit to the oscillation pattern of $B//c'$ at 1.8 K (Fig. 3b, inset), we have obtained quantum mobility ranging from 392 – 1630 cm$^2$/Vs, lower than that in ZrGeS (Table I). In spite of that, we have obtained non-trivial Berry phase for the $F$=37.4T and 360T bands, implying that the topological non-trivial states also exist in ZrGeSe. Table II summarizes all the fitted parameters, including the quantum mobility and Berry phases.

For the case of ZrGeTe, its dHvA oscillations display relatively simpler patterns as compared to ZrGeS and ZrGeSe, without signatures of Zeeman splitting or magnetic breakdown for both $B//c'$ (Figs. 4a-4c) and $B//ab'$ (Figs. 4d-4f). The effective mass for each frequency component obtained from the fit of the temperature dependence of FFT amplitude (insets of Figs. 4c and 4f) ranges from 0.16$m_0$ to 0.24$m_0$, larger than those for ZrGeS and ZrGeSe. The oscillation patterns for both $B//c'$ and $B//ab'$ can be easily fitted to the multiband LK-formula, as shown in the insets of Figs. 4b and 4e, from which we have obtained non-trivial Berry phase for each band (Table III).

From the above analyses, we have revealed evidences of topological fermions in ZrGe*M* (*M*=S, Se, Te), including light effective mass, high quantum mobility, and most importantly, non-trivial Berry phases. These results are consistent with the theoretically predicted topological nodal-line state in *WHM* compounds [46]. From ZrGeS to ZrGeTe, the interlayer binding energy is predicted to become smaller [46] so that we can expect a possible evolution toward 2D in the electronic structure. In order to better understand the dimensionality evolution of the electronic structure in the ZrGe*M* series, we have studied the angular dependence of the quantum oscillations, which can reveal direct information on the Fermi surface morphology.

Using the measurement setup shown in the inset of Fig. 2a, we have measured dHvA oscillations for the ZrGe*M* compounds under different field orientations. After the background subtraction, the oscillation pattern of $\Delta\tau$ displays a clear evolution with the rotation of the magnetic field for each member in ZrGe*M*, as shown in Fig. 5a-5c. In Figs. 5d-5f, we present the angular dependences of the major fundamental frequencies obtained from the FFT analyses. As a cross-check, we have also studied SdH effect in ZrGe*M* through magnetotransport measurements. As shown in Fig. 6, SdH oscillations can be observed for the out-of-plane field orientations ($B//c$). For in-plane fields ($B//ab$), the SdH effect is hardly visible for all three compounds. Furthermore, we have observed negative longitudinal magnetoresistance for ZrGeS, which is possibly associated with chiral anomaly that has been observed in other Dirac materials [1, 73]. The attenuation of the SdH effect with the field rotating toward the in-plane direction has also been observed in ZrSi*M* [50, 52], which seems to be a generic feature for all the *WHM*-type nodal-line semimetals. Nevertheless, in the angle range where SdH oscillations are observable, the extracted SdH oscillation frequencies and their angular dependences agree well with those measured in the dHvA oscillations (Figs. 5d-5f), though some high frequency components are too weak to be measured.

As shown in Figs. 5d-5f, in all the three ZrGe$M$ compounds we find the high-frequency branch is composed of $F_{\beta 1}$ and $F_{\beta 2}$. The difference between $F_{\beta 1}$ and $F_{\beta 2}$ varies remarkably with angle, reaching zero at certain angles, which is suggestive of corrugated cylindrical Fermi surfaces, as widely seen in other quasi-2D systems such as iron-based superconductors [74] and agrees with the layered structure of ZrGe$M$ (Fig. 1a). However, the presence of dHvA oscillations in the whole angle range from $\theta=0°$ to $90°$ clearly indicates 3D Fermi surface. Therefore, the Fermi surface composed of $F_{\beta 1}$/$F_{\beta 2}$ bands should be highly anisotropic. In contrast, the Fermi surface comprised of the $F_\alpha$ band is less anisotropic, which is reflected in the weak angular dependence of $F_\alpha$ (Figs. 5d-5f). The 3D characters of both $F_{\beta 1}$/$F_{\beta 2}$ and $F_\alpha$ bands can be attributed to strong interlayer binding energy in ZrGe$M$ [46]. For instance, the binding energy in ZrGeTe is nearly three times higher than that of ZrSiTe [46].

Given that ZrSi$M$ and ZrGe$M$ are expected to share similar electronic band structures near the Fermi level [46], the comparison between these two families of compounds can provide further information on the nature of the electronic bands probed in the quantum oscillations in ZrGe$M$. As noted above, in the ZrSi$M$ family, ZrSiS has been found to harbor a 3D nodal-line cone state with a very small SOC-induced gap (~20meV), as well as a gapless 2D Dirac cone state protected by the non-symmorphic symmetry [50]. These two Dirac bands are most likely responsible for the high ($F_\beta$) and low ($F_\alpha$) frequency dHvA oscillations respectively [56]. For the ZrGe$M$ compounds, as presented above, our dHvA experiments also reveal high and low frequency oscillation components with non-trivial Berry phase (Tables I-III). The higher frequencies are in the 112-380T range (see Tables I-III), which is of the same order of magnitude as that in ZrSi$M$ (102-240T) [52, 56]. This fact, together with their angular dependences (Figs. 5d-f), suggests that those higher frequency oscillation components most likely originate from the 3D nodal-line Dirac bands. The nature of the lower frequency component ($F_\alpha$), however, is not very clear. The $F_\alpha$ component probed in ZrSiS is attributed to the 2D non-symmorphic Dirac state generated by a Si square-net [56]. The $F_\alpha$ band in ZrGe$M$ might have a similar origin as in ZrSiS, but display a strong 3D character as revealed by the weak angular dependence of $F_\alpha$ (Figs. 5d-f). The 3D nature is likely associated with enhanced interlayer binding in ZrGe$M$. To further clarify the nature of the $F_\alpha$ band in ZrGe$M$, more theoretical and experimental efforts are necessarily needed.

In addition to the dimensionality difference between ZrGe$M$ and ZrSi$M$, the greater ionic radius of Ge is also expected to modify lattice constants, thus affecting the electronic structure [75]. Moreover, the SOC strength is also stronger in ZrGe$M$ due to the fact that Ge atom is heavier than Si atom. Therefore, ZrGe$M$ provides a good platform for examining how these parameters affect the topological fermion properties in the *WHM*-type materials. Previous studies have shown that the lattice constant ratio $c/a$ in *WHM* governs the Dirac node position for the non-symmorphic Dirac cone [53]. The $c/a$ ratio can be tuned by $M$ since the $c$-axis elongates for larger $M$ atoms (Fig. 1a). In ZrSiS which shows a small $c/a \sim 2.27$ [75], the non-symmorphic bands cross below the Fermi level [50], whereas in ZrSiTe, due to the "right" $c/a$ ratio (~2.57), the non-symmorphic Dirac bands cross right at the Fermi level, which allows for investigating the transport properties of the Dirac fermions protected by the non-symmorphic symmetry [53]. Nevertheless, under such a circumstance, the zeroth Landau level of the relativistic fermions is pinned at the Dirac node, and other Landau levels would not pass through the Fermi level upon increasing magnetic field [76], so the quantum oscillations due to the non-symmorphic Dirac fermions were not detected in our previous dHvA experiments on ZrSiTe [52]. In addition to the tuning through $M$ atom, the $c/a$ ratio is also affected by $H$ since the $a$-axis is determined by the square net formed by $H$ atoms (Fig. 1a). Due to the larger ionic radius of Ge, substituting Ge for Si is equivalent to applying lateral tensile strain, thus resulting in even smaller $c/a$ ratios (~ 2.21, 2.23, and 2.22 for $M$=S, Se, and Te, respectively) in ZrGe$M$ than in ZrSiS, so the Dirac node of the non-symmorphic Dirac cone in ZrGe$M$ may be pushed further away from the Fermi level. As a result, the Fermi pocket formed by such Dirac bands is expected to be larger, which is consistent with our experimental observation that $F_\alpha$ probed in ZrGe$M$ (12-62T for $B//c'$) is larger than that the $F_\alpha$ value in ZrSiS (8.4T for $B//c$) [56].

In ZrSiS, the non-symmorphic Dirac cone generated by the Si square lattice does not show a SOC-induced gap due to the protection by the non-symmorphic symmetry, while the nodal-line Dirac cone is gapped by SOC [46, 50], which consequently leads to more massive nodal-line Dirac fermions. In ZrSi$M$, from $M$=S to Se and to Te, the nodal-line Dirac fermions become more massive and quantum mobility become lower [52, 56], which is in line with enhanced SOC-induced gaps due to heavier elements [53]. A similar trend is also observed for the high frequency band (which corresponds to the nodal-line band as discussed above) in ZrGe$M$ (Tables I-III). Furthermore, as noted above, SOC in ZrGe$M$ should be stronger than in

ZrSi*M* for identical *M* atom. Indeed, our experimental results presented above show that for each *M* atom (S, Se or Te), the effective cyclotron mass is heavier and the quantum mobility is lower in ZrGe*M* (Tables I-III) than in ZrSi*M* [52, 56].

In summary, we have synthesized the single crystals of ZrGe*M* (*M*=S, Se and Te) and performed quantum oscillation studies on these materials. The analyses of dHvA quantum oscillation data reveal the experimental evidences for the theoretically predicted topological semimetal state in ZrGe*M* for the first time. The Fermi surfaces comprised of Dirac bands exhibit 3D nature for all the ZrGe*M* compounds, due to enhanced interlayer binding energy. From the comparison between the ZrGe*M* and ZrSi*M* systems, we find that the topological fermion properties can be tuned by lattice constant and SOC through different combination of elements. These findings provide clues to the design and tuning of topological Dirac state in the *WHM*-based materials.

**ACKNOWLEDGMENTS**

This work was supported by the US Department of Energy under grant DE-SC0014208 (support for personnel, material synthesis and magnetization measurements). The work at the National High Magnetic Field Laboratory is supported by the NSF Cooperative Agreement No. DMR-1157490 and the State of Florida (high field magnetic torque and magnetotransport measurements).

**Reference**

[1]  T. Liang, Q. Gibson, M. N. Ali, M. Liu, R. J. Cava and N. P. Ong, Ultrahigh mobility and giant magnetoresistance in the Dirac semimetal $Cd_3As_2$, *Nature Mater.* **14**, 280-284 (2015)

[2]     J. Xiong, S. Kushwaha, J. Krizan, T. Liang, R. J. Cava and N. P. Ong, Anomalous conductivity tensor in the Dirac semimetal $Na_3Bi$, *Europhys. Lett.* **114**, 27002 (2016)

[3]     C. Shekhar, A. K. Nayak, Y. Sun, M. Schmidt, M. Nicklas, I. Leermakers, U. Zeitler, Y. Skourski, J. Wosnitza, Z. Liu, et al., Extremely large magnetoresistance and ultrahigh mobility in the topological Weyl semimetal candidate NbP, *Nature Phys.* **11**, 645-649 (2015)

[4]     N. J. Ghimire, L. Yongkang, M. Neupane, D. J. Williams, E. D. Bauer and F. Ronning, Magnetotransport of single crystalline NbAs, *J. Phys. Condens. Matter* **27**, 152201 (2015)

[5]     Z. Wang, Y. Zheng, Z. Shen, Y. Lu, H. Fang, F. Sheng, Y. Zhou, X. Yang, Y. Li, C. Feng, et al., Helicity-protected ultrahigh mobility Weyl fermions in NbP, *Phys. Rev. B* **93**, 121112 (2016)

[6]     F. Arnold, C. Shekhar, S.-C. Wu, Y. Sun, R. D. dos Reis, N. Kumar, M. Naumann, M. O. Ajeesh, M. Schmidt, A. G. Grushin, et al., Negative magnetoresistance without well-defined chirality in the Weyl semimetal TaP, *Nature Communications* **7**, 11615 (2016)

[7]     X. Huang, L. Zhao, Y. Long, P. Wang, D. Chen, Z. Yang, H. Liang, M. Xue, H. Weng, Z. Fang, et al., Observation of the Chiral-Anomaly-Induced Negative Magnetoresistance in 3D Weyl Semimetal TaAs, *Phys. Rev. X* **5**, 031023 (2015)

[8]     H. Wang, H. Wang, H. Liu, H. Lu, W. Yang, S. Jia, X.-J. Liu, X. C. Xie, J. Wei and J. Wang, Observation of superconductivity induced by a point contact on 3D Dirac semimetal Cd3As2 crystals, *Nature Mater.* **15**, 38-42 (2016)

[9]     Z. Wang, Y. Sun, X.-Q. Chen, C. Franchini, G. Xu, H. Weng, X. Dai and Z. Fang, Dirac semimetal and topological phase transitions in $A_3Bi$(A=Na, K, Rb), *Phys. Rev. B* **85**, 195320 (2012)

[10]    Z. K. Liu, B. Zhou, Y. Zhang, Z. J. Wang, H. M. Weng, D. Prabhakaran, S.-K. Mo, Z. X. Shen, Z. Fang, X. Dai, et al., Discovery of a Three-Dimensional Topological Dirac Semimetal, Na$_3$Bi, *Science* **343**, 864-867 (2014)

[11]    Z. Wang, H. Weng, Q. Wu, X. Dai and Z. Fang, Three-dimensional Dirac semimetal and quantum transport in Cd$_3$As$_2$, *Phys. Rev. B* **88**, 125427 (2013)

[12]    S. Borisenko, Q. Gibson, D. Evtushinsky, V. Zabolotnyy, B. Büchner and R. J. Cava, Experimental Realization of a Three-Dimensional Dirac Semimetal, *Phys. Rev. Lett.* **113**, 027603 (2014)

[13]    Z. K. Liu, J. Jiang, B. Zhou, Z. J. Wang, Y. Zhang, H. M. Weng, D. Prabhakaran, S. K. Mo, H. Peng, P. Dudin, et al., A stable three-dimensional topological Dirac semimetal Cd$_3$As$_2$, *Nature Mater.* **13**, 677-681 (2014)

[14]    M. Neupane, S.-Y. Xu, R. Sankar, N. Alidoust, G. Bian, C. Liu, I. Belopolski, T.-R. Chang, H.-T. Jeng, H. Lin, et al., Observation of a three-dimensional topological Dirac semimetal phase in high-mobility Cd$_3$As$_2$, *Nature Commun.* **5**, 3786 (2014)

[15]    H. Weng, C. Fang, Z. Fang, B. A. Bernevig and X. Dai, Weyl Semimetal Phase in Noncentrosymmetric Transition-Metal Monophosphides, *Phys. Rev. X* **5**, 011029 (2015)

[16]    S.-M. Huang, S.-Y. Xu, I. Belopolski, C.-C. Lee, G. Chang, B. Wang, N. Alidoust, G. Bian, M. Neupane, C. Zhang, et al., A Weyl Fermion semimetal with surface Fermi arcs in the transition metal monopnictide TaAs class, *Nature Commun.* **6**, 7373 (2015)

[17]    S.-Y. Xu, I. Belopolski, N. Alidoust, M. Neupane, G. Bian, C. Zhang, R. Sankar, G. Chang, Z. Yuan, C.-C. Lee, et al., Discovery of a Weyl Fermion semimetal and topological Fermi arcs, *Science* **349**, 613-617 (2015)

[18]   B. Q. Lv, H. M. Weng, B. B. Fu, X. P. Wang, H. Miao, J. Ma, P. Richard, X. C. Huang, L. X. Zhao, G. F. Chen, et al., Experimental Discovery of Weyl Semimetal TaAs, *Phys. Rev. X* **5**, 031013 (2015)

[19]   B. Q. Lv, N. Xu, H. M. Weng, J. Z. Ma, P. Richard, X. C. Huang, L. X. Zhao, G. F. Chen, C. E. Matt, F. Bisti, et al., Observation of Weyl nodes in TaAs, *Nature Physics* **11**, 724–727 (2015)

[20]   L. X. Yang, Z. K. Liu, Y. Sun, H. Peng, H. F. Yang, T. Zhang, B. Zhou, Y. Zhang, Y. F. Guo, M. Rahn, et al., Weyl semimetal phase in the non-centrosymmetric compound TaAs, *Nature Phys.* **11**, 728-732 (2015)

[21]   S.-Y. Xu, N. Alidoust, I. Belopolski, Z. Yuan, G. Bian, T.-R. Chang, H. Zheng, V. N. Strocov, D. S. Sanchez, G. Chang, et al., Discovery of a Weyl fermion state with Fermi arcs in niobium arsenide, *Nature Phys.* **11**, 748-754 (2015)

[22]   N. Xu, H. M. Weng, B. Q. Lv, C. E. Matt, J. Park, F. Bisti, V. N. Strocov, D. Gawryluk, E. Pomjakushina, K. Conder, et al., Observation of Weyl nodes and Fermi arcs in tantalum phosphide, *Nature Commun.* **7**, 11006 (2015)

[23]   L. Lu, Z. Wang, D. Ye, L. Ran, L. Fu, J. D. Joannopoulos and M. Soljačić, Experimental observation of Weyl points, *Science* **349**, 622-624 (2015)

[24]   A. A. Soluyanov, D. Gresch, Z. Wang, Q. Wu, M. Troyer, X. Dai and B. A. Bernevig, Type-II Weyl semimetals, *Nature* **527**, 495-498 (2015)

[25]   Y. Sun, S.-C. Wu, M. N. Ali, C. Felser and B. Yan, Prediction of Weyl semimetal in orthorhombic $MoTe_2$, *Phys. Rev. B* **92**, 161107 (2015)

[26]   F. Y. Bruno, A. Tamai, Q. S. Wu, I. Cucchi, C. Barreteau, A. de la Torre, S. McKeown Walker, S. Riccò, Z. Wang, T. K. Kim, et al., Observation of large topologically trivial Fermi arcs in the candidate type-II Weyl $WTe_2$, *Phys. Rev. B* **94**, 121112 (2016)

[27] C. Wang, Y. Zhang, J. Huang, S. Nie, G. Liu, A. Liang, Y. Zhang, B. Shen, J. Liu, C. Hu, et al., Spectroscopic Evidence of Type II Weyl Semimetal State in WTe$_2$, *arXiv:1604.04218* (2016)

[28] Y. Wu, D. Mou, N. H. Jo, K. Sun, L. Huang, S. L. Bud'ko, P. C. Canfield and A. Kaminski, Observation of Fermi Arcs in Type-II Weyl Semimetal Candidate WTe$_2$, *Phys. Rev. B* **94**, 121113(R) (2016)

[29] K. Deng, G. Wan, P. Deng, K. Zhang, S. Ding, E. Wang, M. Yan, H. Huang, H. Zhang, Z. Xu, et al., Experimental observation of topological Fermi arcs in type-II Weyl semimetal MoTe$_2$, *arXiv:1603.08508* (2016)

[30] L. Huang, T. M. McCormick, M. Ochi, Z. Zhao, M.-t. Suzuki, R. Arita, Y. Wu, D. Mou, H. Cao, J. Yan, et al., Spectroscopic evidence for type II Weyl semimetal state in MoTe$_2$, *arXiv:1603.06482* (2016)

[31] J. Jiang, Z. K. Liu, Y. Sun, H. F. Yang, R. Rajamathi, Y. P. Qi, L. X. Yang, C. Chen, H. Peng, C.-C. Hwang, et al., Observation of the Type-II Weyl Semimetal Phase in MoTe2, *arXiv:1604.00139* (2016)

[32] A. Liang, J. Huang, S. Nie, Y. Ding, Q. Gao, C. Hu, S. He, Y. Zhang, C. Wang, B. Shen, et al., Electronic Evidence for Type II Weyl Semimetal State in MoTe2, *arXiv:1604.01706* (2016)

[33] N. Xu, Z. J. Wang, A. P. Weber, A. Magrez, P. Bugnon, H. Berger, C. E. Matt, J. Z. Ma and B. Q. L. B. B. Fu, N. C. Plumb, M. Radovic, E. Pomjakushina, K. Conder, T. Qian, J. H. Dil, J. Mesot, H. Ding, M. Shi, Discovery of Weyl semimetal state violating Lorentz invariance in MoTe$_2$, *arXiv:1604.02116* (2016)

[34] S. Borisenko, D. Evtushinsky, Q. Gibson, A. Yaresko, T. Kim, M. N. Ali, B. Buechner, M. Hoesch and R. J. Cava, Time-Reversal Symmetry Breaking Type-II Weyl State in YbMnBi2, *arXiv:1507.04847* (2015)

[35]  G. Chang, S.-Y. Xu, H. Zheng, B. Singh, C.-H. Hsu, I. Belopolski, D. S. Sanchez, G. Bian, N. Alidoust, H. Lin, et al., Room-temperature magnetic topological Weyl fermion and nodal line semimetal states in half-metallic Heusler Co$_2$TiX (X=Si, Ge, or Sn), *Sci. Rep.* **6**, 38839 (2016)

[36]  Z. Wang, M. G. Vergniory, S. Kushwaha, M. Hirschberger, E. V. Chulkov, A. Ernst, N. P. Ong, R. J. Cava and B. A. Bernevig, Time-Reversal Breaking Weyl Fermions In Magnetic Heuslers, *arXiv:1603.00479* (2016)

[37]  J. Kübler and C. Felser, Weyl points in the ferromagnetic Heusler compound Co$_2$MnAl, *EPL (Europhysics Letters)* **114**, 47005 (2016)

[38]  H. Yang, Y. Sun, Y. Zhang, W.-J. Shi, S. S. P. Parkin and B. Yan, Topological Weyl semimetals in the chiral antiferromagnetic materials Mn$_3$Ge and Mn$_3$Sn, *arXiv:1608.03404* (2016)

[39]  Y. Zhang, Y. Sun, H. Yang, J. Železný, S. P. P. Parkin, C. Felser and B. Yan, Strong, anisotropic anomalous Hall effect and spin Hall effect in chiral antiferromagnetic compounds Mn3X (X = Ge, Sn, Ga, Ir, Rh and Pt), *arXiv:1610.04034* (2016)

[40]  G. Chang, B. Singh, S.-Y. Xu, G. Bian, S.-M. Huang, C.-H. Hsu, I. Belopolski, N. Alidoust, D. S. Sanchez, H. Zheng, et al., Theoretical prediction of magnetic and noncentrosymmetric Weyl fermion semimetal states in the R-Al-X family of compounds (R=rare earth, Al=aluminium, X=Si, Ge), *arXiv:1604.02124* (2016)

[41]  Y. Kim, B. J. Wieder, C. L. Kane and A. M. Rappe, Dirac Line Nodes in Inversion-Symmetric Crystals, *Phys. Rev. Lett.* **115**, 036806 (2015)

[42]  H. Weng, Y. Liang, Q. Xu, R. Yu, Z. Fang, X. Dai and Y. Kawazoe, Topological node-line semimetal in three-dimensional graphene networks, *Phys. Rev. B* **92**, 045108 (2015)

[43]  L. S. Xie, L. M. Schoop, E. M. Seibel, Q. D. Gibson, W. Xie and R. J. Cava, A new form of Ca$_3$P$_2$ with a ring of Dirac nodes, *APL Mater.* **3**, 083602 (2015)

[44]   R. Yu, H. Weng, Z. Fang, X. Dai and X. Hu, Topological Node-Line Semimetal and Dirac Semimetal State in Antiperovskite $Cu_3PdN$, *Phys. Rev. Lett.* **115**, 036807 (2015)

[45]   M. Zeng, C. Fang, G. Chang, Y.-A. Chen, T. Hsieh, A. Bansil, H. Lin and L. Fu, Topological semimetals and topological insulators in rare earth monopnictides, *arXiv:1504.03492* (2015)

[46]   Q. Xu, Z. Song, S. Nie, H. Weng, Z. Fang and X. Dai, Two-dimensional oxide topological insulator with iron-pnictide superconductor LiFeAs structure, *Phys. Rev. B* **92**, 205310 (2015)

[47]   H. Huang, S. Zhou and W. Duan, Type-II Dirac fermions in the $PtSe_2$ class of transition metal dichalcogenides, *Phys. Rev. B* **94**, 121117 (2016)

[48]   G. Bian, T.-R. Chang, R. Sankar, S.-Y. Xu, H. Zheng, T. Neupert, C.-K. Chiu, S.-M. Huang, G. Chang, I. Belopolski, et al., Topological nodal-line fermions in spin-orbit metal $PbTaSe_2$, *Nature Commun.* **7**, 10556 (2016)

[49]   G. Bian, T.-R. Chang, H. Zheng, S. Velury, S.-Y. Xu, T. Neupert, C.-K. Chiu, S.-M. Huang, D. S. Sanchez, I. Belopolski, et al., Drumhead Surface States and Topological Nodal-Line Fermions in $TlTaSe_2$, *Phys. Rev. B* **93**, 121113 (2016)

[50]   L. M. Schoop, M. N. Ali, C. Straszer, A. Topp, A. Varykhalov, D. Marchenko, V. Duppel, S. S. P. Parkin, B. V. Lotsch and C. R. Ast, Dirac cone protected by non-symmorphic symmetry and three-dimensional Dirac line node in ZrSiS, *Nature Commun.* **7**, 11696 (2016)

[51]   M. Neupane, I. Belopolski, M. M. Hosen, D. S. Sanchez, R. Sankar, M. Szlawska, S.-Y. Xu, K. Dimitri, N. Dhakal, P. Maldonado, et al., Observation of Topological Nodal Fermion Semimetal Phase in ZrSiS, *Phys. Rev. B* **93**, 201104 (2016)

[52]  J. Hu, Z. Tang, J. Liu, X. Liu, Y. Zhu, D. Graf, K. Myhro, S. Tran, C. N. Lau, J. Wei, et al., Evidence of Topological Nodal-Line Fermions in ZrSiSe and ZrSiTe, *Phys. Rev. Lett.* **117**, 016602 (2016)

[53]  T. Andreas, M. L. Judith, V. Andrei, D. Viola, V. L. Bettina, R. A. Christian and M. S. Leslie, Non-symmorphic band degeneracy at the Fermi level in ZrSiTe, *New J. Phys.* **18**, 125014 (2016)

[54]  M. M. Hosen, K. Dimitri, I. Belopolski, P. Maldonado, R. Sankar, N. D. G. D. T. Cole, P. M. Oppeneer, D. Kaczorowski, F. Chou, M. Z. Hasan, et al., Tunability of the topological nodal-line semimetal phase in ZrSiX-type materials, *arXiv:1612.07018*

[55]  Y. Wu, L.-L. Wang, E. Mun, D. D. Johnson, D. Mou, L. Huang, Y. Lee, S. L. Bud/'ko, P. C. Canfield and A. Kaminski, Dirac node arcs in PtSn4, *Nature Phys.* **12**, 667-671 (2016)

[56]  J. Hu, Z. Tang, J. Liu, Y. Zhu, J. Wei and Z. Mao, Evidence of Dirac cones with 3D character probed by dHvA oscillations in nodal-line semimetal ZrSiS, *arXiv:1604.01567* (2016)

[57]  R. Singha, A. Pariari, B. Satpati and P. Mandal, Titanic magnetoresistance and signature of non-degenerate Dirac nodes in ZrSiS, *arXiv:1602.01993* (2016)

[58]  M. N. Ali, L. M. Schoop, C. Garg, J. M. Lippmann, E. Lara, B. Lotsch and S. Parkin, Butterfly Magnetoresistance, Quasi-2D Dirac Fermi Surfaces, and a Topological Phase Transition in ZrSiS, *arXiv:1603.09318* (2016)

[59]  X. Wang, X. Pan, M. Gao, J. Yu, J. Jiang, J. Zhang, H. Zuo, M. Zhang, Z. Wei, W. Niu, et al., Evidence of Both Surface and Bulk Dirac Bands and Anisotropic Nonsaturating Magnetoresistance in ZrSiS, *Advanced Electronic Materials* **2**, 1600228 (2016)

[60]   M. M. Hosen, K. Dimitri, I. Belopolski, P. Maldonado, R. Sankar, N. Dhakal, G. Dhakal, T. Cole, P. M. Oppeneer, D. Kaczorowski, et al., Tunability of the topological nodal-line semimetal phase in ZrSiX-type materials, *arXiv:1612.07018* (2016)

[61]   R. Lou, J. Z. Ma, Q. N. Xu, B. B. Fu, L. Y. Kong, Y. G. Shi, P. Richard, H. M. Weng, Z. Fang, S. S. Sun, et al., Emergence of topological bands on the surface of ZrSnTe crystal, *Phys. Rev. B* **93**, 241104 (2016)

[62]   D. Takane, Z. Wang, S. Souma, K. Nakayama, C. X. Trang, T. Sato, T. Takahashi and Y. Ando, Dirac-node arc in the topological line-node semimetal HfSiS, *Phys. Rev. B* **94**, 121108 (2016)

[63]   K. M. Nitesh Kumar, Yanpeng Qi, Shu-Chun Wu, Lei Wang, Binghai Yan, Claudia Felser, Chandra Shekhar, Unusual magneto-transport from Si-square nets in topological semimetal HfSiS, *arXiv:1612.05176* (2016)

[64]   X. X. C. Chen, J. Jiang, S. -C. Wu, Y. P. Qi, L. X. Yang, M. X. Wang, Y. Sun, N.B.M. Schröter, H. F. Yang, L. M. Schoop, Y. Y. Lv, J. Zhou, Y. B. Chen, S. H. Yao, M. H. Lu, Y. F. Chen, C. Felser, B. H. Yan, Z. K. Liu, Y. L. Chen, Dirac Line-nodes and Effect of Spin-orbit Coupling in Nonsymmorphic Critical Semimetal MSiS (M=Hf, Zr), *arXiv:1701.06888* (2017)

[65]   H. Onken, K. Vierheilig and H. Hahn, Über Silicid- und Germanidchalkogenide des Zirkons und Hafniums, *Z. Anorg. Allg. Chem.* **333**, 267-279 (1964)

[66]   A. J. Klein Haneveld and F. Jellinek, Zirconium silicide and germanide chalcogenides preparation and crystal structures, *Recl. Trav. Chim. Pays-Bas* **83**, 776-783 (1964)

[67]   I. M. Lifshitz and A. M. Kosevich, Theory of Magnetic Susceptibility in Metals at Low Temperatures, *Sov. Phys. JETP* **2**, 636-645 (1956)

[68]   D. Shoenberg, *Magnetic Oscillations in Metals*. Cambridge Univ. Press: Cambridge, 1984.

[69]     G. P. Mikitik and Y. V. Sharlai, Manifestation of Berry's Phase in Metal Physics, *Phys. Rev. Lett.* **82**, 2147-2150 (1999)

[70]     J. Hu, J. Y. Liu, D. Graf, S. M. A. Radmanesh, D. J. Adams, A. Chuang, Y. Wang, I. Chiorescu, J. Wei, L. Spinu, et al., π Berry phase and Zeeman splitting of Weyl semimetal TaP, *Sci. Rep.* **6**, 18674 (2016)

[71]     M. V. Kartsovnik, High Magnetic Fields: A Tool for Studying Electronic Properties of Layered Organic Metals, *Chem. Rev.* **104**, 5737-5782 (2004)

[72]     S. G. Sharapov, V. P. Gusynin and H. Beck, Magnetic oscillations in planar systems with the Dirac-like spectrum of quasiparticle excitations, *Phys. Rev. B* **69**, 075104 (2004)

[73]     J. Xiong, S. K. Kushwaha, T. Liang, J. W. Krizan, M. Hirschberger, W. Wang, R. J. Cava and N. P. Ong, Evidence for the chiral anomaly in the Dirac semimetal Na3Bi, *Science* **350**, 413-416 (2015)

[74]     A. Carrington, Quantum oscillation studies of the Fermi surface of iron-pnictide superconductors, *Rep. Prog. Phys.* **74**, 124507 (2011)

[75]     C. Wang and T. Hughbanks, Main Group Element Size and Substitution Effects on the Structural Dimensionality of Zirconium Tellurides of the ZrSiS Type, *Inorg. Chem.* **34**, 5524-5529 (1995)

[76]     J. Y. Liu, J. Hu, D. Graf, T. Zou, M. Zhu, Y. Shi, S. Che, S. M. A. Radmanesh, C. N. Lau, L. Spinu, et al., Unusual interlayer quantum transport behavior caused by the zeroth Landau level in YbMnBi$_2$, *arXiv:1608.05956* (2016)

**Table I**. Parameters derived from the analyses of dHvA oscillations for ZrGeS. $F$, oscillation frequency; $T_D$, Dingle temperature; $m^*$, effective mass; $\tau$, quantum relaxation time; $\mu_q$, quantum mobility; $\phi_B$, Berry phase.

|  | $F$ (T) | $m^*/m_0$ | $T_D$ (K) | $\tau$ (ps) | $\mu$ (cm$^2$/Vs) | $\phi_B$ |
|---|---|---|---|---|---|---|
|  | 12.5 | 0.05 | - | - | - | - |
| $B//c'$ | 236 | 0.062 | - | - | - | - |
|  | 380 | 0.097 | - | - | - | - |
| $B//ab'$ | 17 | 0.05 | 28 | 0.044 | 1547 | $(-0.38\pm0.25)\pi$ |
|  | 132 | 0.09 | 35 | 0.035 | 684 | $(-0.22\pm0.25)\pi$ |

**Table II**. Parameters derived from the analyses of dHvA oscillations for ZrGeSe. $F$, oscillation frequency; $T_D$, Dingle temperature; $m^*$, effective mass; $\tau$, quantum relaxation time; $\mu_q$, quantum mobility; $\phi_B$, Berry phase.

|  | $F$ (T) | $m^*/m_0$ | $T_D$ (K) | $\tau$ (ps) | $\mu$ (cm$^2$/Vs) | $\phi_B$ |
|---|---|---|---|---|---|---|
|  | 37.4 | 0.11 | 12 | 0.102 | 1630 | $(0.78\pm0.25)\pi$ |
| $B//c'$ | 226 | 0.22 | 25 | 0.049 | 392 | $(-0.28\pm0.25)\pi$ |
|  | 360 | 0.12 | 35 | 0.035 | 513 | $(1.21\pm0.25)\pi$ |
|  | 17.3 | 0.05 | - | - | - | - |
| $B//ab'$ | 112 | 0.13 | - | - | - | - |
|  | 167 | 0.17 | - | - | - | - |

**Table III**. Parameters derived from the analyses of dHvA oscillations for ZrGeTe. $F$, oscillation frequency; $T_D$, Dingle temperature; $m^*$, effective mass; $\tau$, quantum relaxation time; $\mu_q$, quantum mobility; $\phi_B$, Berry phase.

|  | $F$ (T) | $m^*/m_0$ | $T_D$ (K) | $\tau$ (ps) | $\mu$ (cm$^2$/Vs) | $\phi_B$ |
|---|---|---|---|---|---|---|
| $B//c'$ | 62 | 0.16 | 30 | 0.041 | 451 | $(1.20\pm0.25)\pi$ |
|  | 265 | 0.20 | 27 | 0.046 | 404 | $(0.44\pm0.25)\pi$ |
| $B//ab'$ | 13 | 0.21 | 14.4 | 0.086 | 720 | $(-1.08\pm0.25)\pi$ |
|  | 156 | 0.23 | 14.6 | 0.084 | 640 | $(0.60\pm0.25)\pi$ |
|  | 203 | 0.24 | 16 | 0.077 | 564 | $(0.70\pm0.25)\pi$ |

# Figures

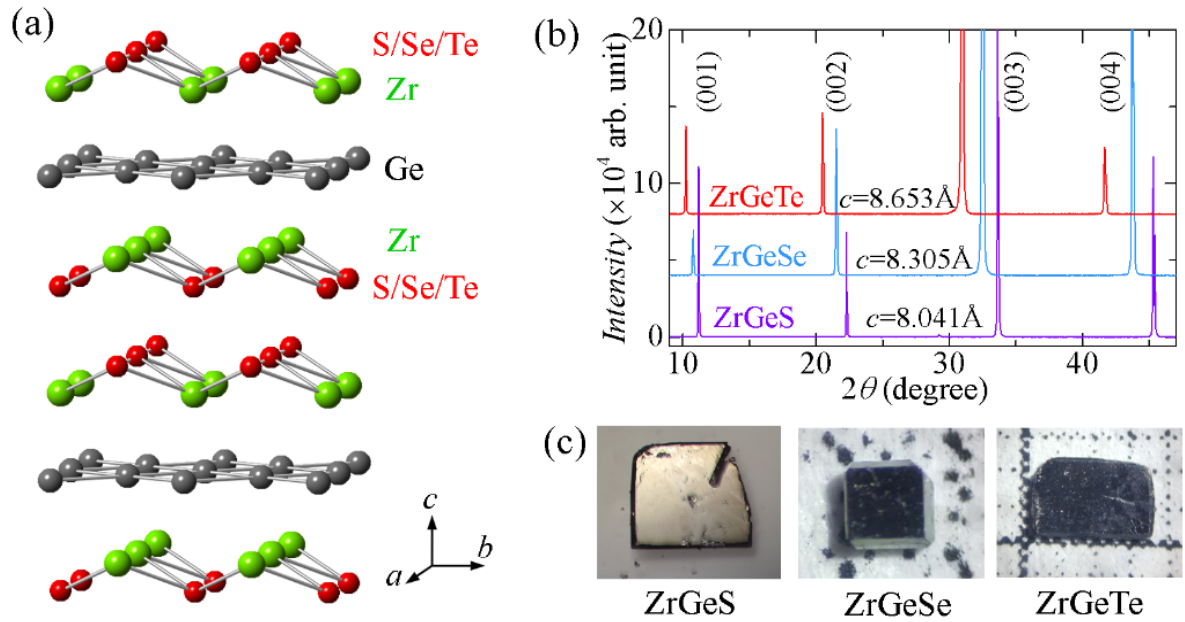

**FIG. 1.** (a) Crystal structure of ZrGe*M* (*M*=S, Se, Te). (b) Single crystal x-ray diffraction patterns for ZrGe*M*. (c) Images of ZrGe*M* single crystals.

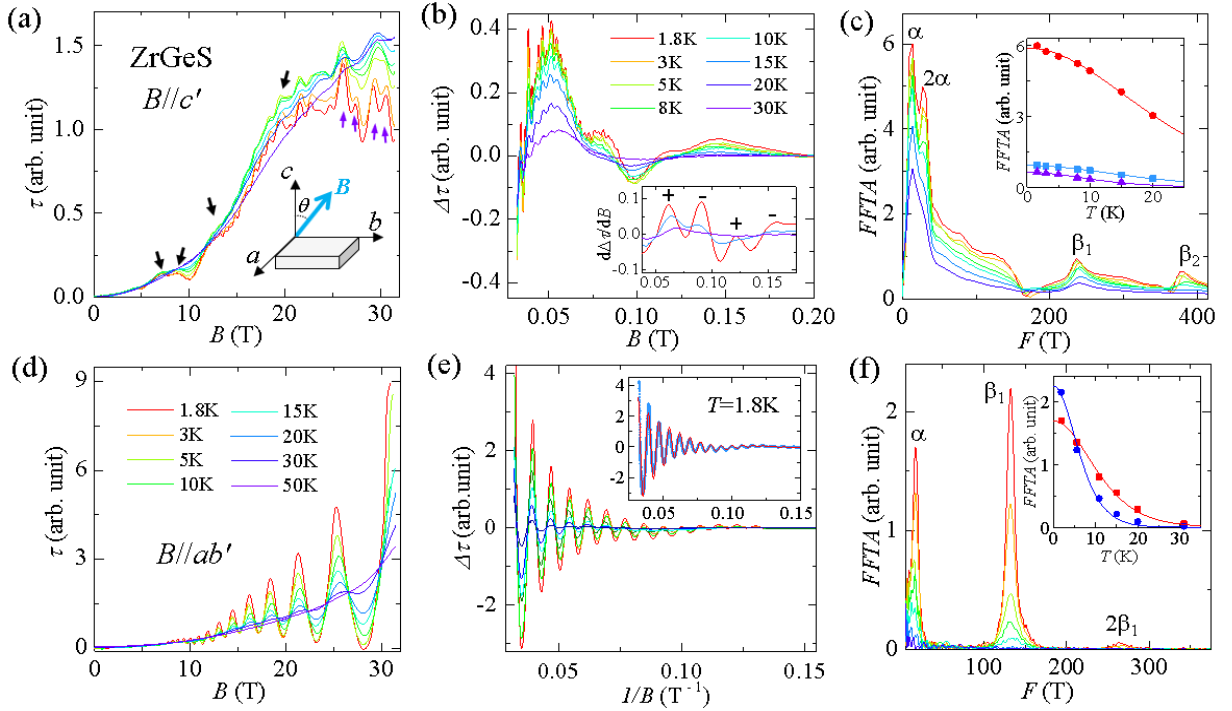

**FIG. 2.** (a) The field dependence of magnetic torque $\tau$ for ZrGeS at different temperatures, which displays strong dHvA oscillations. The magnetic field is applied along the nearly out-of-plane direction ($B//c'$). The black and purple arrows indicate the splitting of oscillation peaks for the $F_\alpha$ (=12.5T) and $F_{\beta1}$ (=236T) components. Inset: the experiment setup. (b) The oscillatory component of $\tau$ for $B//c'$. Inset: Zeeman splitting in susceptibility oscillations at $T$=1.8, 15, and 30K. (c) The FFT spectra for the oscillatory component $\Delta\tau$ for $B//c'$. Inset: the temperature dependence of the FFT amplitude of the major fundamental frequencies, and the fits to the LK formula (solid lines). (d) The field dependence of magnetic torque $\tau$ for ZrGeS at different temperatures. The magnetic field is applied along the nearly in-plane direction ($B//ab'$). (e) The oscillatory component of $\tau$ for $B//ab'$. Inset: the fit of the oscillation pattern at $T$=1.8K to the multi-band LK formula. (f) The FFT spectra for the oscillatory component $\Delta\tau$ for $B//c'$. Inset: the temperature dependence of the FFT amplitude of the major fundamental frequencies, and the fit to the LK formula (solid lines).

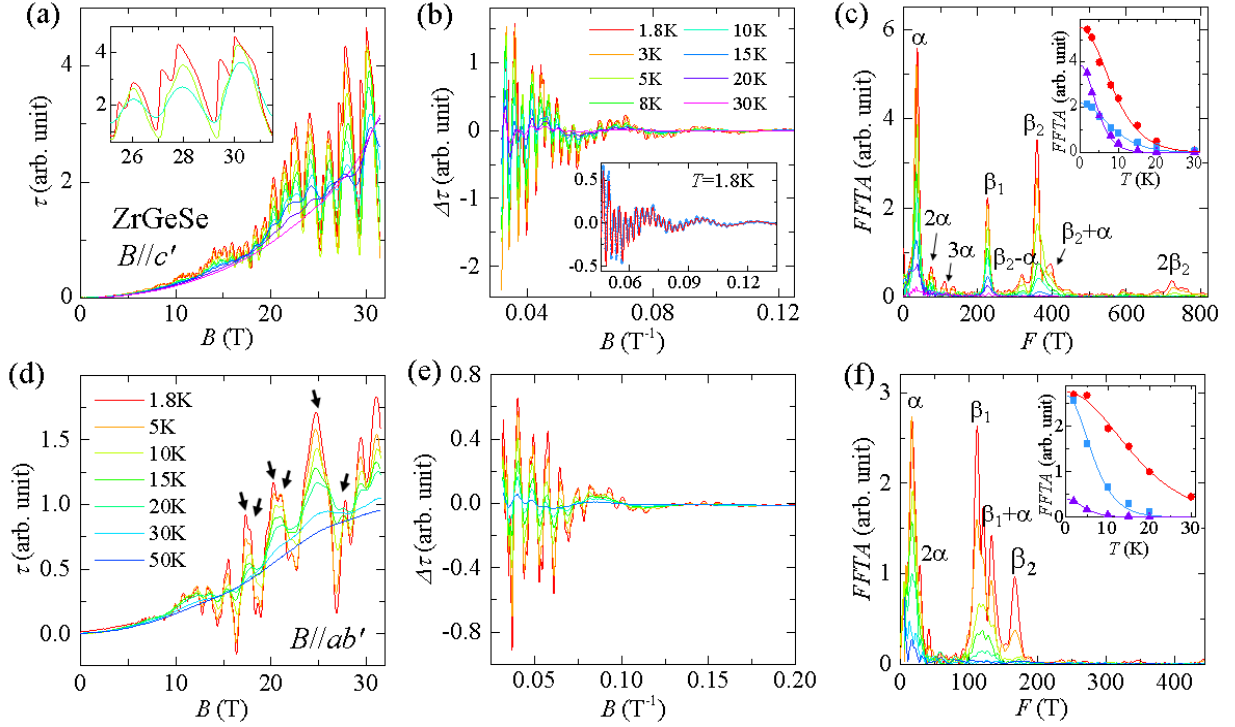

**FIG. 3.** (a) The field dependence of magnetic torque $\tau$ for ZrGeSe at different temperatures for $B//c'$. Inset: dHvA oscillations in $\tau$ under high field, which displays inverse saw-tooth pattern and Zeeman splitting. (b) The oscillatory component of $\tau$ for $B//c'$. Inset: the fit of the oscillation pattern at $T$=1.8K to the multi-band LK formula. (c) The FFT spectra for the oscillatory component $\Delta\tau$ for $B//c'$. Inset: the temperature dependence of the FFT amplitude of the major fundamental frequencies, and the fits to the LK formula (solid lines). (d) The field dependence of magnetic torque $\tau$ for ZrGeSe at different temperatures for $B//ab'$. The black arrows indicate split peaks. (e) The oscillatory component of $\tau$ for $B//ab'$. (f) The FFT spectra for the oscillatory component $\Delta\tau$ for $B//c'$. Inset: the temperature dependence of the FFT amplitude of the major fundamental frequencies, and the fits to the LK formula (solid lines).

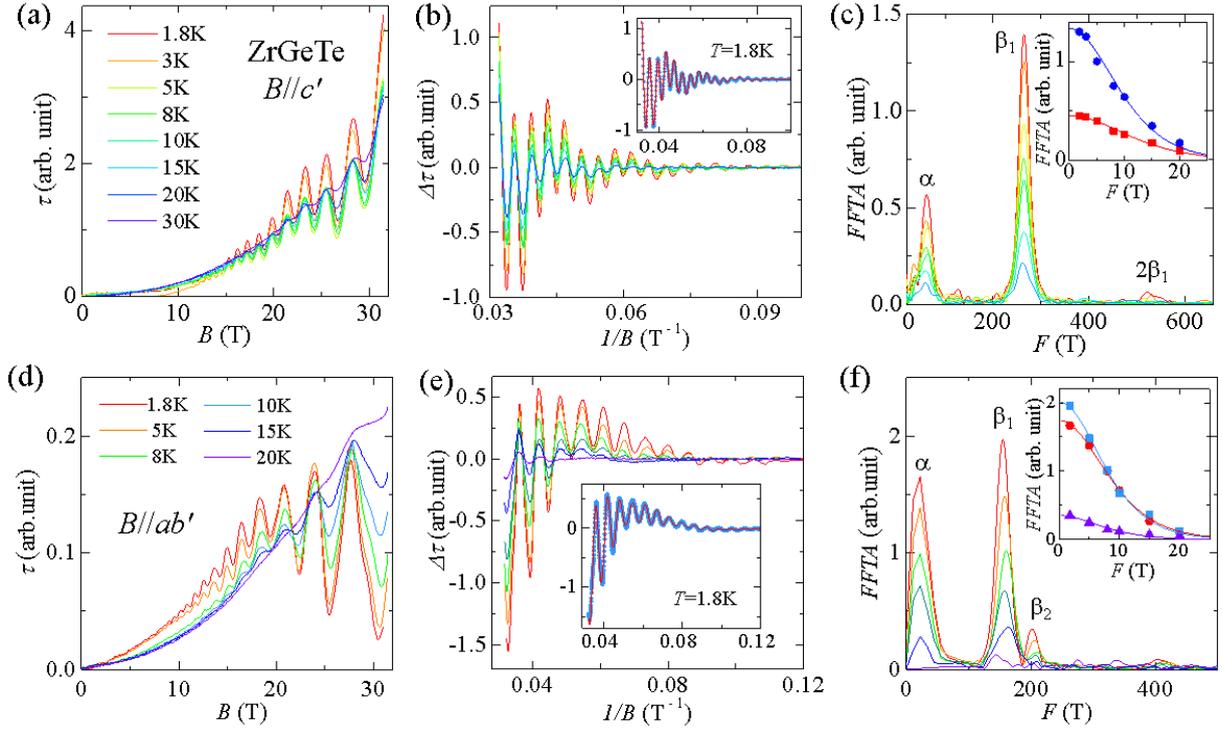

**FIG. 4.** (a) The field dependence of magnetic torque $\tau$ for ZrGeTe at different temperatures for $B//c'$. (b) The oscillatory component of $\tau$ for $B//c'$. Inset: the fit of the oscillation pattern at $T$=1.8K to the multi-band LK formula. (c) The FFT spectra for the oscillatory component $\Delta\tau$ for $B//c'$. Inset: the temperature dependence of the FFT amplitude of the major fundamental frequencies, and the fits to the LK formula (solid lines). (d) The field dependence of magnetic torque $\tau$ for ZrGeTe at different temperatures for $B//ab'$. (e) The oscillatory component of $\tau$ for $B//ab'$. Inset: the fit of the oscillation pattern at $T$=1.8K to the multi-band LK formula. (f) The FFT spectra for the oscillatory component $\Delta\tau$ for $B//c'$. Inset: the temperature dependence of the FFT amplitude of the major fundamental frequencies, and the fits to the LK formula (solid lines).

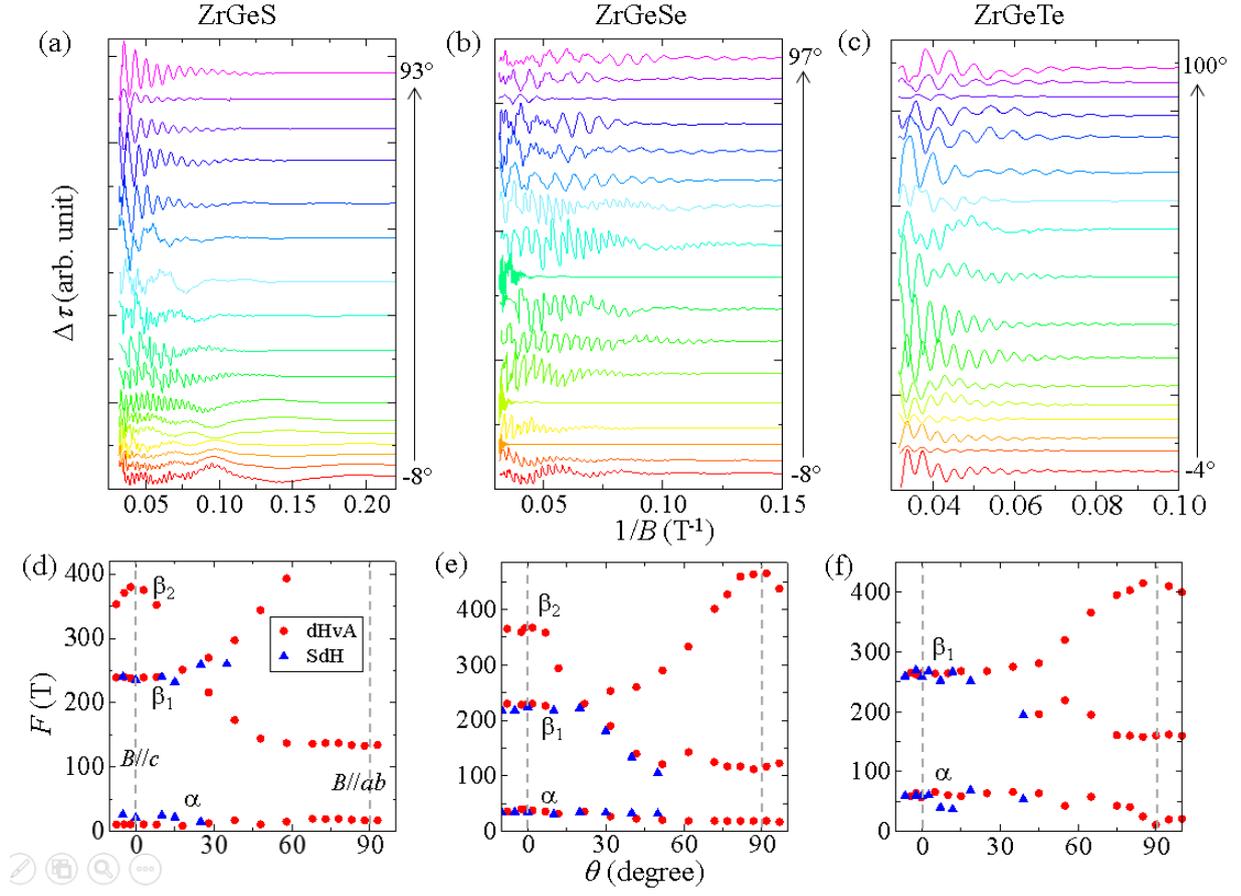

**FIG. 5.** (a-c) dHvA oscillations of (a) ZrGeS, (b) ZrGeSe, and (c) ZrGeTe at $T$=1.8K under different magnetic field orientations. The data collected under at different $\theta$ have been shifted for clarity. (d-f) The angular dependence of oscillation frequencies for (d) ZrGeS, (e) ZrGeSe, and (f) ZrGeTe. The gray dashed lines indicate the in-plane ($B//c$, 0 degree) and out-of-plane ($B//ab$, 90 degree) directions.

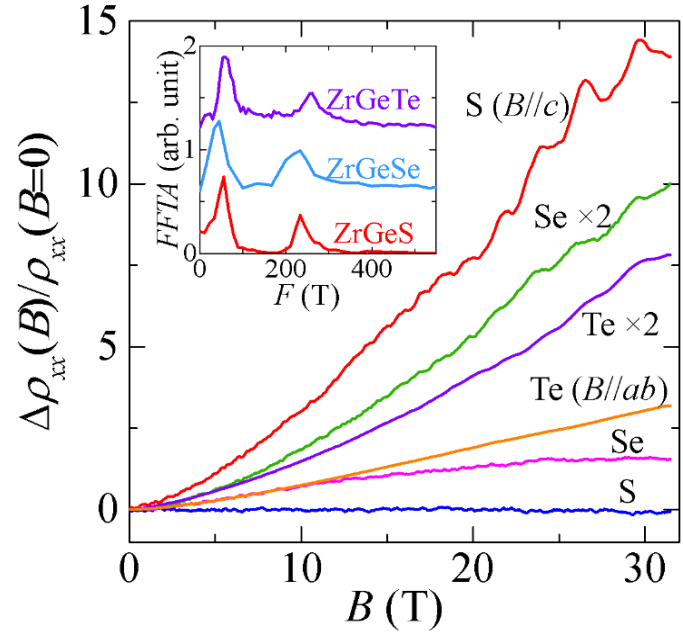

**FIG. 6**. In-plane magnetoresistance and SdH oscillations of ZrGe*M* (*M*=S, Se, Te) for *B//c* and *B//ab*. The data for ZrGeSe and ZrGeTe for *B//c* have been multiplied by a factor of 2 for clarity. Inset: The FFT spectra for the SdH oscillations; the data have been shifted for clarity.